\documentclass[rsi,
 amsmath,amssymb,
 reprint,%
]{revtex4-1}

\usepackage{graphicx}% Include figure files
\usepackage{dcolumn}% Align table columns on decimal point
\usepackage{bm}% bold math
\usepackage[rgb,table]{xcolor}

\usepackage[utf8]{inputenc}
\usepackage[T1]{fontenc}
\usepackage{mathptmx}
\usepackage{textcomp}

\usepackage{siunitx}

\begin{document}

\preprint{AIP/123-QED}

\title[Measurements of the energy distribution of electrons lost from the minimum B-field - the effect of instabilities and two-frequency heating]{Measurements of the energy distribution of electrons lost from the minimum B-field - the effect of instabilities and two-frequency heating}
% Force line breaks with \\

\author{I. Izotov$^1$}
 \email{ivizot@ipfran.ru}
\author{O. Tarvainen$^2$}%
\author{V. Skalyga$^{1,3}$}
\author{D. Mansfeld$^1$}
\author{H. Koivisto$^4$}%
\author{R. Kronholm$^4$}
\author{V Toivanen$^4$}
\author{V. Mironov$^5$}

 \affiliation{$^1$Institute of Applied Physics of Russian Academy of Sciences, 603950, Nizhny Novgorod, Russia.\\
 $^2$STFC, ISIS Pulsed Spallation Neutron and Muon Facility, Rutherford Appleton Laboratory, Harwell OX11 0QX, United Kingdom.\\
 $^3$Lobachevsky State University of Nizhny Novgorod, 603950, Nizhny Novgorod, Russia.\\
 $^4$University of Jyväskylä, 40500 Jyväskylä, Finland.\\
 $^5$Joint Institute for Nuclear Research, 141980, Dubna, Russia.}% 

\date{\today}% It is always \today, today,

\begin{abstract}
Further progress in the development of ECR ion sources (ECRIS) requires deeper understanding of the underlying physics. One of the topics that remains obscure, though being crucial for the performance of the ECRIS, is the electron energy distribution (EED). A well-developed technique of measuring the EED of electrons escaping axially from the magnetically confined plasma of an ECRIS was used for the study of EED in unstable mode of plasma confinement, i.e. in the presence of kinetic instabilities. The experimental data were recorded for pulsed and CW discharges with a room-temperature 14 GHz ECRIS at the JYFL accelerator laboratory. The measurements were focused on observing differences between the EED escaping from a stable and unstable plasmas. It was found that nonlinear phenomena alter the EED noticeably. The electron losses are enhanced in both unstable regime and with two-frequency heating suppressing the instabilities. It has been shown earlier that two-frequency heating boosts the ECRIS performance presumably owing to the suppression of instabilities. We report the observed changes in EED introduced by the secondary frequency in different regimes, including an off-resonance condition where the secondary frequency is lower than the minimum frequency satisfying the resonance condition for cold electrons at the magnetic field minimum. Finally, we suggest an experimental method of qualitative evaluation of the energy distribution of electrons confined in the magnetic trap using a method of measuring energy distribution of lost electrons during the plasma decay in pulsed operation of the ion source.
\end{abstract}

\maketitle

\section{Introduction}

Electron cyclotron resonance ion sources (ECRIS) have been essential in fundamental nuclear physics research and applications over the past several decades. They are used in a wide range of accelerator facilities for the production of highly charged ion beams of stable and radioactive elements, from hydrogen up to uranium. The ECRIS performance crucially depends on highly charged ions trapping efficiency. The ion confinement time is a complex function of the ion temperature $Ti$ and electrostatic potential dip $\Delta\Phi$, caused by the accumulation of magnetically confined electrons to the core plasma and different loss rates of electrons and ions resulting in an ambipolar potential barrier restricting the ion losses \cite{Melin, Pastukhov}. Improvements of the magnetic plasma confinement over the past years have yielded significant enhancements of the ECRIS performance. However, despite the fact that the enhancement of the magnetic systems leads to an obvious increase of the ion confinement time, high charge state ion production is impossible if the mean electron energy is too low. Thus, knowledge on electron energy distribution (EED) is essential for tuning the ECRIS for the best performance in terms of high charge state production. Measurement of the plasma and wall bremsstrahlung is a traditional technique of estimating the EED of an ECRIS plasma. However, such measurement gives information only on a so-called spectral temperature (i.e. slope of a linear fit to bremsstrahlung spectrum in log scale) and maximum electron energy. Despite being suitable for relative evaluation of the electron heating efficiency, these parameters do not allow to estimate the ionization rate of a particular charge state. Thus, knowledge on the EED is essential for further improvement of ECRIS performance. Furthermore, it has been shown that the ECRIS plasma is prone to kinetic instabilities \cite{Olli_beam}, which leads to a dramatic decrease in performance \cite{Olli_limitations_invited, Olli_limitations}, i.e. in the high charge state current. The onset of kinetic instabilities is determined by the EED of confined high-energy electrons \cite{Izotov_PSST_2015, Golubev_PhysRev, Melin_proc}, which highlights the necessity of gathering information on the EED to understand the underlying mechanism and to mitigate the instabilities \cite{Skalyga_2freq_stabilization}.
The present work is an overview of the recent experiments devoted to direct measurements of the lost electrons energy distribution (LEED), i.e. the energy distribution of electrons escaping axially the magnetic confinement of conventional minimum-B ECRIS. The data reported here focuses on the temporal dynamics of the LEED and transition between stable and unstable (i.e. in the presence of kinetic instabilities) regimes. CW and pulsed operation with single and double frequency heating were investigated. Also, some relevant results obtained in stable mode \cite{Izotov_EEDF_2018}  are presented. It is emphasized that the EED of the confined electrons in the magnetic trap and the LEED are very likely different. However, it may be argued that the LEED reflects the EED of the confined electrons at least in terms of parametric dependencies.

\section{Experimental setup}
The experimental data were taken with the JYFL 14 GHz ECRIS. The source uses a permanent magnet sextupole and two solenoid coils. The superposition of the solenoid and sextupole fields forms a minimum-B structure for plasma confinement. The maximum strength of the permanent magnet sextupole (i.e. the radial field) is 1.09 T at the chamber wall. The axial field strength can be varied by adjusting the solenoid currents, which affects the injection and extraction mirror ratios as well as the $B_{min}/B_{ECR}$ ratio ($B_{ECR}=0.5$ T at 14 GHz). The solenoid field configuration is best described by the values at injection ($B_{inj}$), minimum ($B_{min}$), and extraction ($B_{ext}$). Typically used settings, corresponding to a normal ECRIS operation, are $B_{min}/B_{ECR}=0.75$, the values are $B_{inj}=1.976$ T, $B_{min}=0.375$ T, and $B_{ext}=0.913$ T. $B_{min}/B_{ECR}$ ratio is given later for each experiment being the most convenient for describing the magnetic field strength. The solenoid currents were kept identical and  were adjusted simultaneously. Typical operating neutral gas pressures are in the $10^{-7}$ mbar range and the plasma electrons are heated by 100–600 W of microwave power at 14 GHz. The source is equipped with a secondary waveguide port connected to a 10.75–13.75 GHz TWT amplifier with 350–400 W maximum power used simultaneously with 14 GHz microwaves for two-frequency heating experiments. The secondary heating frequency is launched through a magic tee equipped with a microwave diode and connected to an  oscilloscope with attenuator and power limiter for detecting the electromagnetic plasma emission in unstable mode. This signal is used as a triggering signal during experiments in unstable mode. Bismuth germanate (BGO) scintillator coupled with a current-mode photomultiplier tube (Na doped CsI) was used as a (relatively) fast x-ray detector placed outside the plasma chamber near a radial diagnostics port. These complementary diagnostics yield information on the hot electrons interacting with the plasma EM-wave and emitting microwave radiation, wall bremsstrahlung (power) flux induced by energetic electrons escaping the confinement.

\begin{figure}
%/media/NAS_public/EXPERIMENTS/JYFL2018/2018.01.24/Power sweep/plotAll.m
\includegraphics[width=\linewidth]{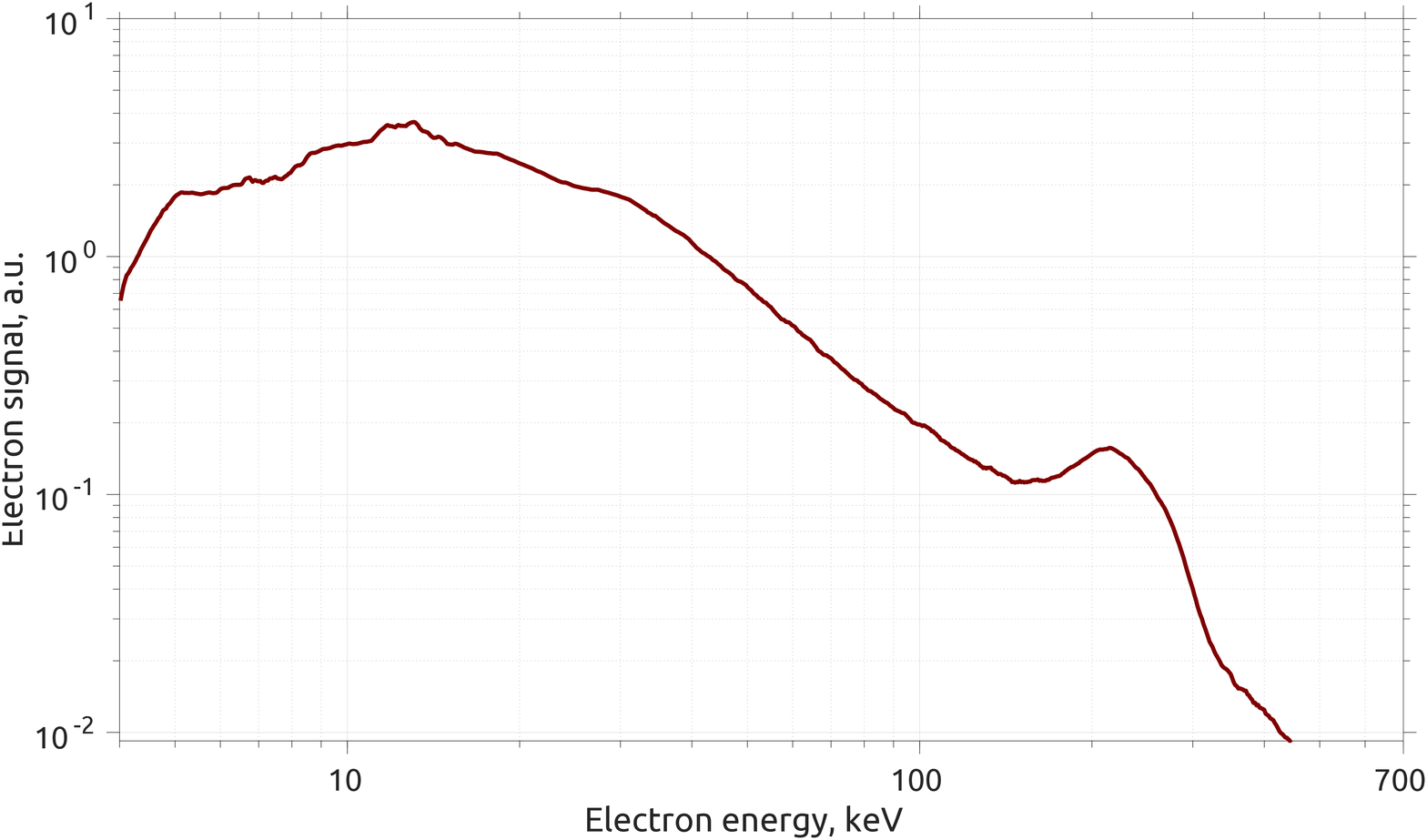}
\caption{\label{fig:fig1} An example of lost electron energy distribution obtained with 600 W microwave power at 14 GHz, $B_{min}/B_{ECR}=0.75$ and $3.5\cdot10^{-7}$ mbar of oxygen.}
\end{figure}

The electrons escaping the confinement were detected with a secondary electron amplifier placed in the beamline downstream from the 90\textdegree ~bending magnet used as an energy dispersive separator. The electron flux was limited by two f=5 mm collimators placed between the ion source and the bending magnet and yet another f=5 mm entrance collimator in front of the secondary electron amplifier. All collimators are made of aluminium and grounded. The polarity of the bending magnet power supply was changed from the normal operation where the magnet is used for $m/q$-separation of high charge state positive ions. The magnetic field deflecting the electrons was measured with a calibrated Hall-probe. The energy distribution of the electrons escaping from the confinement was then determined by ramping the field of the bending magnet and detecting the electron current from the amplifier with either a picoammeter or fast transimpedance amplifier. The energy resolution of the setup provided by the set of collimators and the distance between them  is estimated to be better than 500 eV. The energy dependent transmission efficiency of the electrons leaking from the ion source through the beamline sections and the bending magnet was calculated assuming that the electron distribution at the extraction aperture is independent of energy and has a KV-distribution \cite{Toivanen_KV}. The first two collimators sample a fraction of the beam, which is directly proportional to the energy of the electron beam as long as the beam completely illuminates the collimators (electron energy <100 MeV). Furthermore, the energy dependent yield \cite{SEY} of the secondary electrons released from the amplifier cathode was taken into account during the data analysis together with electron back-scattering coefficient \cite{BSC}. The power supply used for operating the bending magnet coil had a high precision and small current step with the maximum current limited to a value corresponding to the electron energy of 800 keV. The amplifier functions by emitting secondary electrons from biased aluminium cathode and multiplying their number by a chain of subsequent grid stages before measuring the current from the grounded anode. The cathode of the secondary electron amplifier was biased negatively to -4 kV with respect to the laboratory ground, thus prohibiting the detection of electrons with energies below 4 keV. More details together with the experimental scheme may be found in Ref.~\cite{Izotov_EEDF_2018}. The lost electrons energy distribution (LEED) was measured as a function of the ion source parameters e.g. microwave power, microwave frequency (or frequencies) and (axial) magnetic field in different modes of operation (CW/pulsed, single/double frequency heating, stable/unstable plasma). The plasma chamber of the ion source and all focusing electrodes were grounded throughout the experiment, meaning that the detected electron flux consists of the electrons leaking from the plasma through the extraction aperture retarded only by the plasma potential of approximately 20 V \cite{Tarvainen_2005}.
Typical LEED obtained in the stable regime using the technique described above with heating power of 600 W at 14 GHz, $B_{min}/B_{ECR}=0.75$ and $3.5\cdot10^{-7}$ mbar of oxygen is shown in Figure~\ref{fig:fig1}. The average energy in Figure~\ref{fig:fig1} is equal to 65 keV, estimated as $\langle \varepsilon \rangle = L^{-1}\int_{\varepsilon_{min}}^{\varepsilon_{max}} f(\varepsilon) \varepsilon d\varepsilon$, where $L=\int_{\varepsilon_{min}}^{\varepsilon_{max}} f(\varepsilon) d\varepsilon$ measures the total electron losses. A distinct feature of the LEED is the high-energy hump, visible at $\sim$200~keV energy. The hump is correlated with the $B_{min}$ absolute value and, apparently, appears as a result of interaction of electrons with low-frequency (several GHz) electromagnetic wave of yet unknown origin. The hump contains 10-15\% of the total electron flux and accounts for more than 30\% of measured energy losses, which makes it of fundamental interest.

\section{Temporal evolution of the LEED during single instability event}
\begin{figure}[h]
%/media/NAS_public/EXPERIMENTS/JYFL2018/2018.01.26/scan5/plotEEDF_normalized.m
\includegraphics[width=\linewidth]{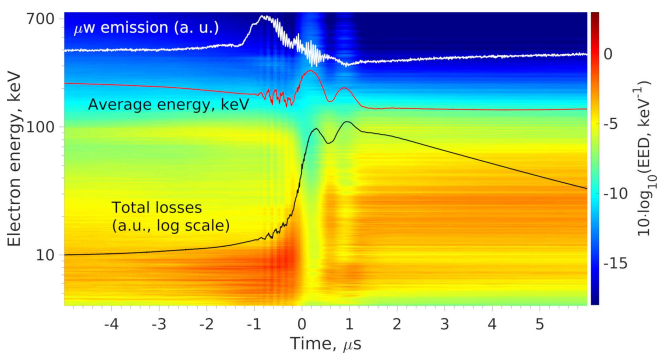}
\caption{\label{fig:fig2}  LEED evolution within a single instability event. Source settings: single frequency heating at 14 GHz, 400 W, $B_{min}/B_{ECR}=0.842$, $3.5 \cdot 10^{-7}$ mbar oxygen pressure.}
\end{figure}

A typical LEED evolution within several microseconds before and after a single instability event (in a periodic instability onset regime) is shown in Figure~\ref{fig:fig2}. The ion source settings were the following: single frequency heating at 14 GHz, 400 W, $B_{min}/B_{ECR}=0.842$, $3.5 \cdot 10^{-7}$ mbar oxygen pressure, corresponding to an unstable plasma. 
The data shown in Figure~\ref{fig:fig2} was assembled from a set of waveforms recorded for each energy bin, triggered by an instability event (microwave emission detected with a microwave diode). The LEED magnitude shown in false color and log scale is normalized to unity square within each time bin. Logarithmic scale is needed as the signal level within the instability burst itself overcomes the background signal by more than 3 orders of magnitude.

The temporal evolution of the microwave emission (white line, arbitrary units), total electron losses (black line, arbitrary units, log scale) and average energy of electrons (red line, same scale as LEED) may be divided into 5 subsequent regions. The first region, i.e. $t < -2~\SI{}{\micro\second}$, corresponds to a quasi-stationary stage, when the LEED changes slowly: the average energy gradually decreases approaching the instability event, whereas total losses increase. The second region, namely $-1.2<t<-0.2 ~\SI{}{\micro\second}$, is where the microwave emission reaches its maximum power. Electron losses are growing fast, and periodic oscillations are visible in all signals, indicating a strong non-stationary process being presumably the interaction of electrons with the electromagnetic wave. The third region, $0<t<1.2~\SI{}{\micro\second}$, corresponds to the stage when the axial electron losses reach their maximum. Two consequent ``bursts” of electrons are seen within this stage. We would like to draw Reader's attention to the notable delay of $\sim$1~$\SI{}{\micro\second}$ between the microwave emission maximum and the first burst of electrons. The delay cannot be explained by the electron time of flight, which is on the order of nanoseconds, but is rather affected by the change of (hot) electron confinement during the onset of the instability. The fourth region at $t>3 ~\SI{}{\micro\second}$ features a gradual decrease of the total losses and an increase of the average energy, gradually merging with the first stage leading to the next instability event.

\begin{figure}[h]
%/media/NAS_public/EXPERIMENTS/JYFL2018/2018.01.26/scan5/plotEEDFs_2D.m
\includegraphics[width=\linewidth]{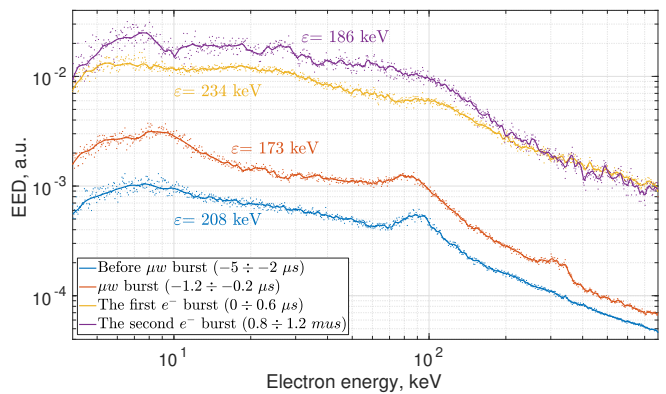}
\caption{\label{fig:fig3}LEED at different instances during an instability onset. Source settings: same as in Fig. \ref{fig:fig2}.}
\end{figure}

Figure~\ref{fig:fig3}(b) shows selected LEEDs: before the instability onset (averaged for $-5<t<-2~ \SI{}{\micro\second}$, blue curve), during the microwave pulse (averaged for $-1.2<t<-0.2 ~ \SI{}{\micro\second}$, red curve), during the first (averaged for $0<t<0.6 ~ \SI{}{\micro\second}$, yellow curve) and the second (averaged for $0.8<t<1.2 ~ \SI{}{\micro\second}$) electron bursts. Despite the obvious difference in the number of expelled electrons (being the area under the curves), the most pronounced differences are the high-energy humps: a single hump at $\sim$100 keV before the microwave pulse, two humps during the microwave pulse ($\sim$100 and $\sim$300 keV) and no humps during the electron bursts. The hump at higher energy is presumably linked to the $B_{min}$ value as discussed in the previous section. The hump at lower energy could be of similar origin, as a rich electromagnetic spectrum interacting with the electrons is generated during the instability event \cite{Izotov_PSST_2015, Mansfeld_PPCF_2016, Izotov_PoP_2017}.

\section{The effect of power modulation on the LEED and plasma stability}

\begin{figure}[h]
%/media/NAS_public/EXPERIMENTS/JYFL2019/2019.02.12/scan2/plotEEDF.m
\includegraphics[width=\linewidth]{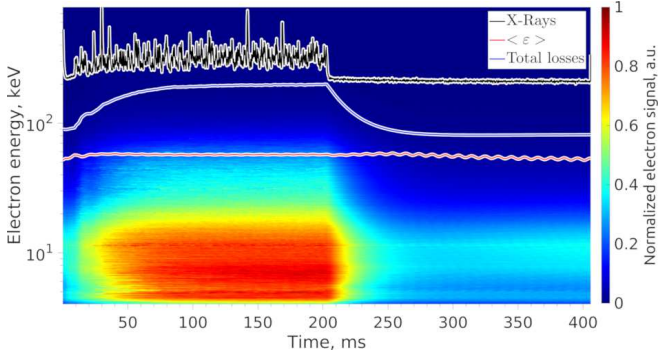}
\caption{\label{fig:fig4} LEED evolution during the transition between unstable and stable plasma. Source settings: single frequency heating at 12 GHz, 192 W (200 ms) / 77 W (200 ms), $B_{min}/B_{ECR}=0.92$, $3.5 \cdot 10^{-7}$ mbar oxygen pressure .}
\end{figure}

It has been shown \cite{Olli_beam} that the heating power is the second most important parameter affecting the plasma stability (the most important being the magnetic field). Figure~\ref{fig:fig4} shows the temporal evolution of the LEED, acquired under periodic power modulation from 192 W (200 ms) to 77 W (200 ms) at 12 GHz, resulting in stable plasma at low power and unstable plasma at high power. The source parameters were the following: $B_{min}/B_{ECR}=0.92$ and $3.5 \cdot 10^{-7}$ mbar oxygen pressure. The curves in Figure~\ref{fig:fig4} are the bremsstrahlung flux (black, arbitrary units), the total electron flux (blue, arbitrary units) and the average energy (red, keV). The x-rays signal (black) clearly shows periodic bursts of x-rays in the high power regime where the plasma is unstable.

\begin{figure}
%/media/NAS_public/EXPERIMENTS/JYFL2019/2019.02.12/scan2/plotEEDFs.m
\includegraphics[width=\linewidth]{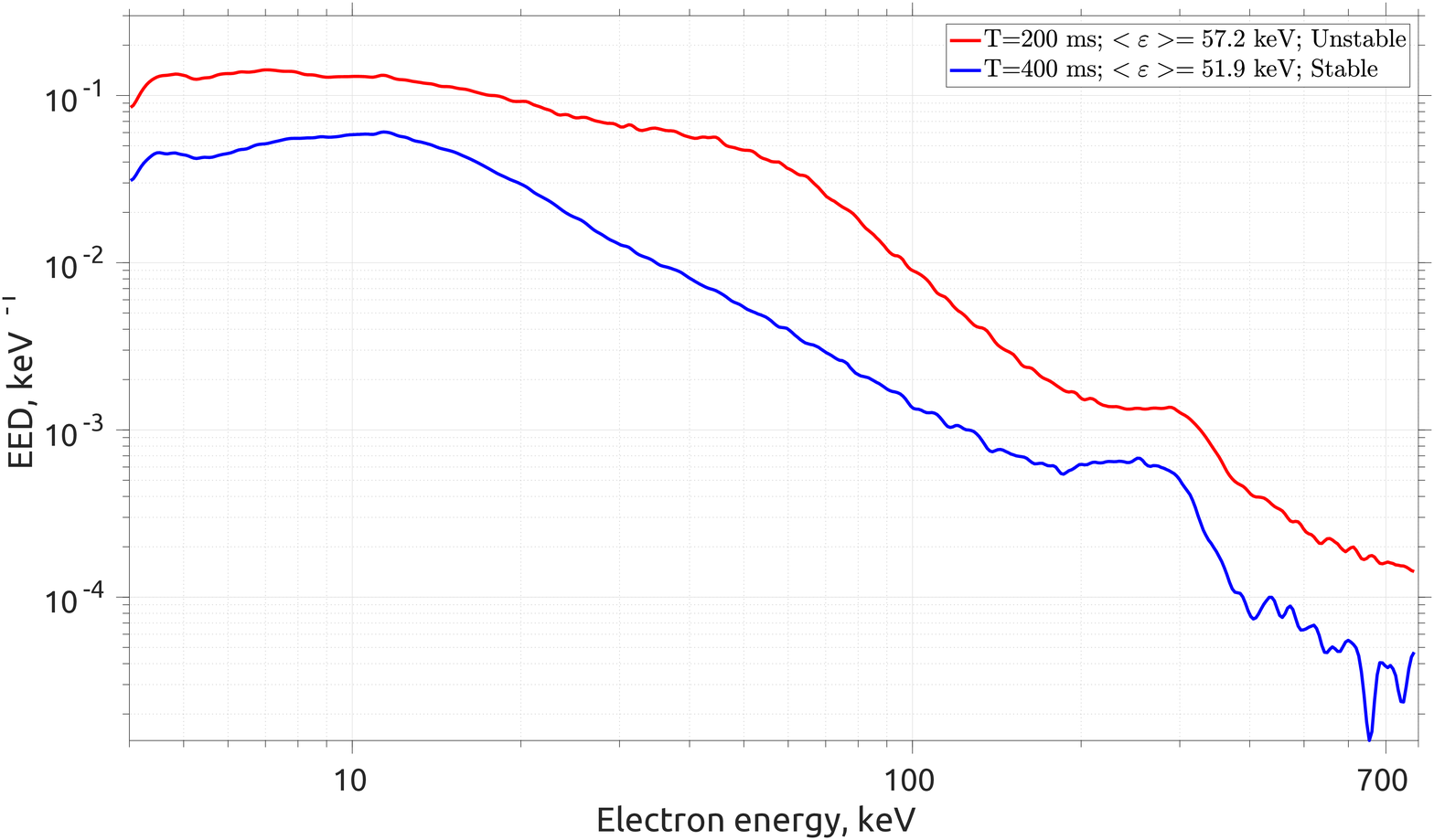}
\caption{\label{fig:fig5} LEED in stable and unstable plasma. Source settings: same as in Fig.~\ref{fig:fig4}.}
\end{figure}

The total electron flux is notably higher in the unstable mode compared to the stable one. This is consistent with observations showing that instabilities expel a large  amount of electric charge \cite{Tarvainen_RSI_2016}. Although the average energy does not change much, the shape of the LEED changes noticeably. This is shown in Figure~\ref{fig:fig5}, where the red curve represents the LEED at t=200 ms, i.e. in the unstable regime, and the blue curve - at t=400 ms, i.e. in stable regime, both taken well after transients between the two regimes. Besides the total flux, the LEED in unstable mode shows enhanced losses in the energy range of $20<\varepsilon<100$ keV when compared to the stable regime, which are the electrons supposedly driving the cyclotron instabilities and reaching the loss cone at the given energy range \cite{Shalashov_EPL_2018}. The high-energy hump is present in both LEEDs, though it is more pronounced in the stable mode, which is consistent with the current understanding of the rf-scattering process causing the hump due to interaction with a low-frequency em-wave.

\section{Power modulation: transition to CW microwave emission regime}

\begin{figure}
%/media/NAS_public/EXPERIMENTS/JYFL2019/2019.02.12/scan3/plotEEDF.m
\includegraphics[width=\linewidth]{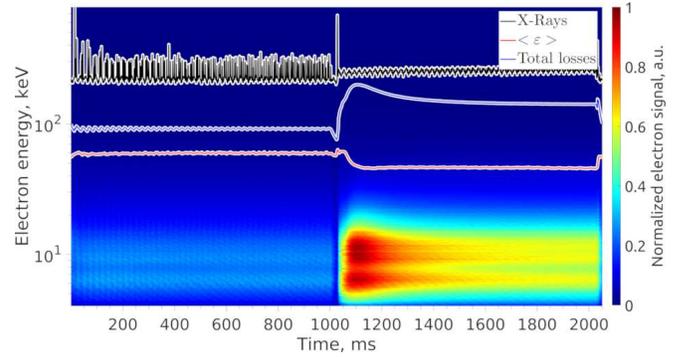}
\caption{\label{fig:fig6}  LEED evolution during the transition between unstable and CW emission plasma. Source settings: single frequency heating at 12 GHz, 15 W (1000 ms) / 50 W (1000 ms), $B_{min}/B_{ECR}=0.98$, $3.5 \cdot 10^{-7}$ mbar oxygen pressure.}
\end{figure}

\begin{figure}
%/media/NAS_public/EXPERIMENTS/JYFL2019/2019.02.12/scan3/plotEEDFs.m
\includegraphics[width=\linewidth]{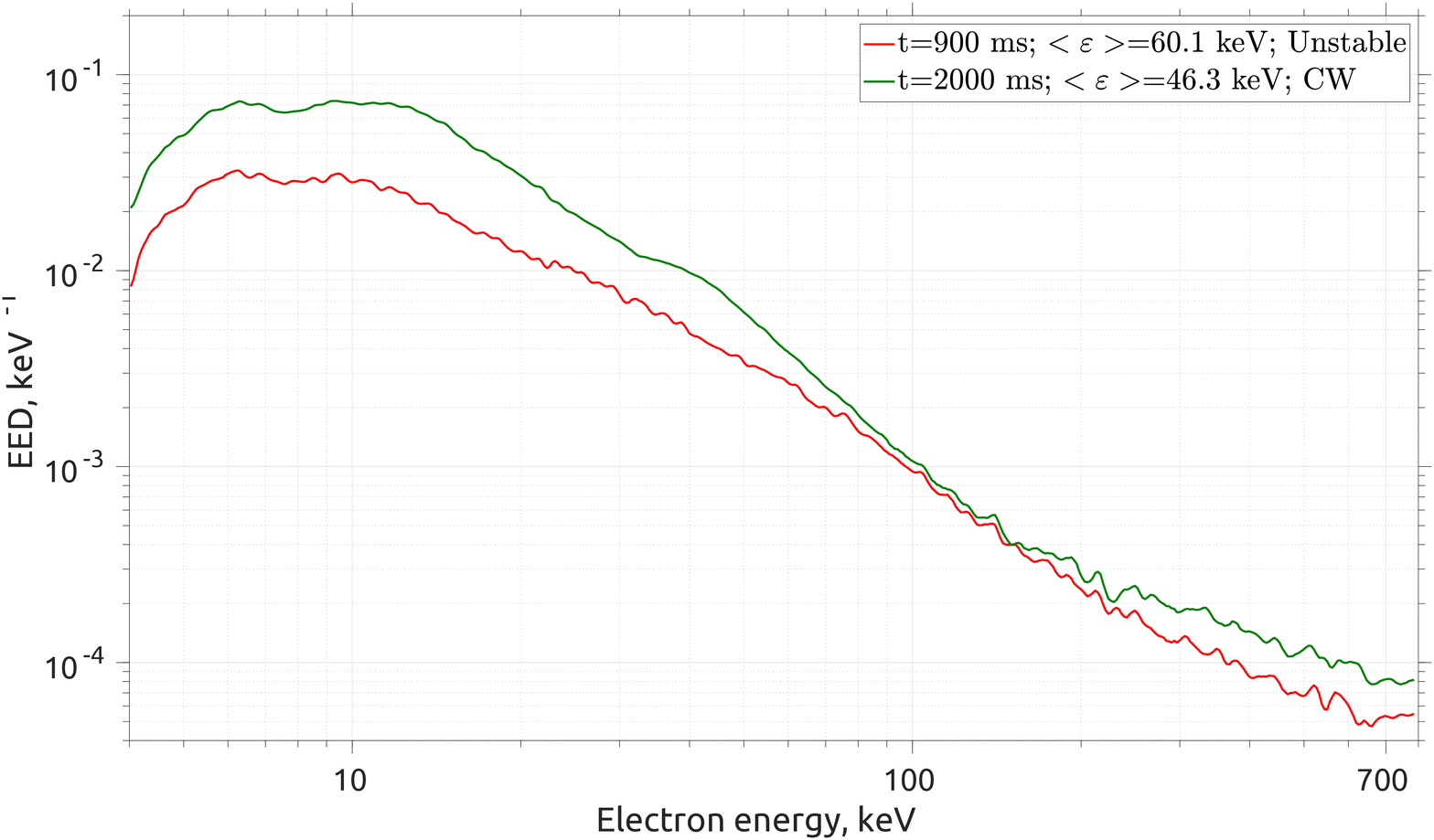}
\caption{\label{fig:fig7} LEED in unstable and CW emission plasma. Source settings: same as in Fig.~\ref{fig:fig6}.}
\end{figure}

It is possible to shift from the unstable regime of plasma confinement to a quasi-stable one, featured by CW electromagnetic emission instead of pulse-periodic one, by further increase of the heating power under certain source settings as described elsewhere \cite{Shalashov_PRL_2018}. Figure~\ref{fig:fig6} shows the temporal evolution of the LEED, acquired under periodic power modulation from 15 W (1000 ms) to 50 W (1000 ms) at 12 GHz, resulting in unstable plasma at low power and quasi-stable (CW regime) plasma at high power. The source parameters were the following: $B_{min}/B_{ECR}=0.98$ and $3.5 \cdot 10^{-7}$ mbar oxygen pressure. The CW regime is characterized by a noticeably higher total electron flux when compared to the unstable regime (i.e. significantly higher than in the stable regime, see Figure~\ref{fig:fig5}). The red curve in Figure~\ref{fig:fig7} represents the LEED at t=900 ms, i.e. in the unstable regime, and the green curve -- at t=2000 ms, i.e. in the CW regime, both taken well after transients between the two regimes. The losses in the CW regime are much greater at energies below 80 keV and slightly greater above 200 keV than in the unstable regime, whereas they are similar in between these energies. This means that the average energy is lower in the unstable regime when compared to the CW one, which is clearly seen in Figure~\ref{fig:fig7}. The CW regime is typically found at source settings well above the instability thresholds \cite{Olli_beam, Shalashov_PRL_2018}, in particular, at very high $B_{min}/B_{ECR}$ ratio.

\section{Plasma stabilization by two-frequency heating}

\begin{figure}
%/media/NAS_public/EXPERIMENTS/JYFL2019/2019.02.11/scan1/plotEEDF.m
\includegraphics[width=\linewidth]{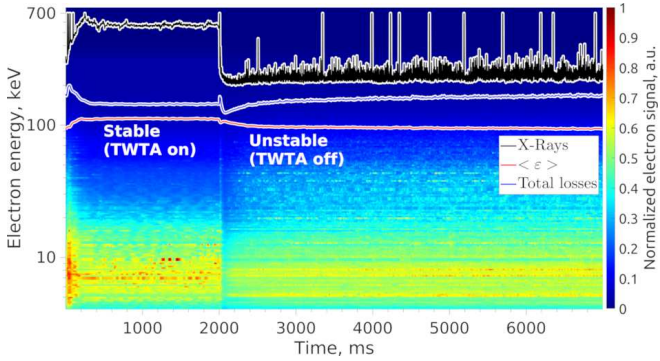}
\caption{\label{fig:fig8} LEED evolution during the transition between stable (double frequency heating) and unstable (single frequency heating) plasma. Source settings: 300 W at 14 GHz + 150 W at 12.7 GHz (2 s) / 300 W at 14 GHz (5 s), $B_{min}/B_{ECR}=0.83/0.92$ (calculated for 14 GHz and 12.7 GHz respectively) and $3.5 \cdot 10^{-7}$ mbar oxygen pressure.}
\end{figure}

Two-frequency heating was first introduced in the 1990s \cite{2freq} and is widely used in modern ECRISs. The essence of the method lies in the injection of lower frequency (and usually lower power) microwaves, in addition to the primary microwave radiation, into the ECRIS plasma. Such technique allows increasing the average charge and current of the extracted ion beam in comparison to single frequency heating (even) at the same level of total injected microwave power. The method affects especially those charge states higher than the peak of the extracted charge state distribution. Two-frequency heating has yet another significant and positive effect on the characteristics of the extracted ion beams. It has been reported \cite{Vondrasek_RSI_2006, Kitagawa_ECRIS12} that introducing the secondary frequency can significantly increase the microwave power range in which the ion current is stable. Thus, the method allows ECRISs to operate properly at settings above the instability threshold observed in single-frequency operation, i.e. higher microwave power and magnetic field strength and, therefore, produce higher currents of highly charged ions as predicted by the semi-empirical scaling law \cite{Geller_scaling}. The root-cause of stabilizing effect of the second frequency has been discussed in \cite{Skalyga_2freq_stabilization}. 
Figure~\ref{fig:fig8} shows the temporal evolution of the LEED, acquired under periodic power modulation from 150 W (2 s) to 0 W (7 s) at 12.7 GHz on top of constant power of 300 W at 14 GHz, resulting in stable plasma in two-frequency heating regime and unstable plasma in 14 GHz-only regime. The source parameters were the following: $B_{min}/B_{ECR}=0.83/0.92$ (calculated for 14 GHz and 12.7 GHz respectively) and $3.5 \cdot 10^{-7}$ mbar oxygen pressure.

\begin{figure}
%/media/NAS_public/EXPERIMENTS/JYFL2019/2019.02.11/scan1/plotEEDFs.m
\includegraphics[width=\linewidth]{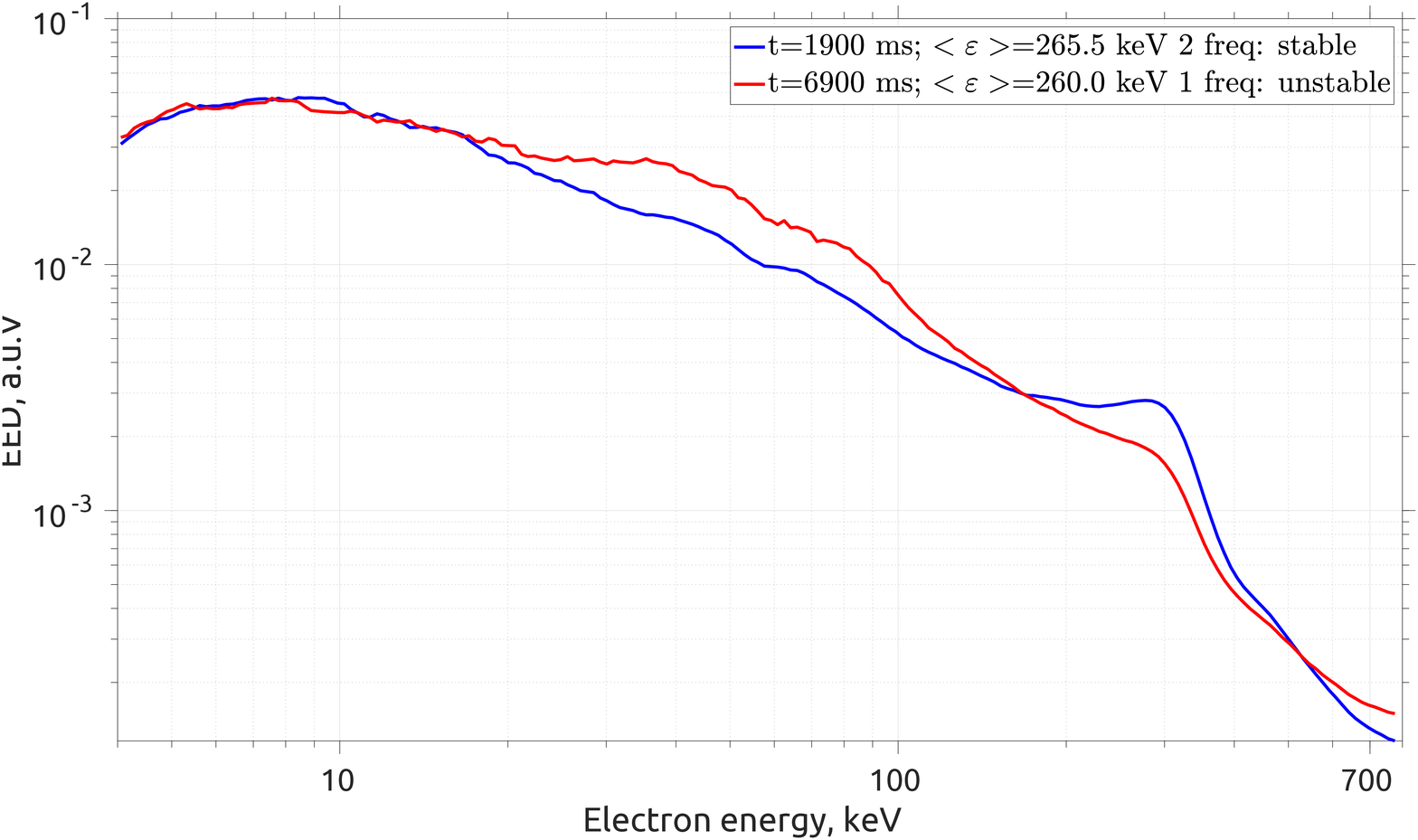}
\caption{\label{fig:fig9} LEED in stabilized by 2-frequency heating and unstable plasma. Source settings: same as in Fig.~\ref{fig:fig6}}
\end{figure}

The total losses in stable (two-frequency regime) and unstable (single frequency regime) are almost identical, which differs significantly from the case of power modulation, when the losses were considerably different in stable and unstable regimes (see Figure~\ref{fig:fig5}). Similar to Figure~\ref{fig:fig5}, the difference in LEED’s shape (see Figure~\ref{fig:fig9}) is pronounced in the energy range of $20<\varepsilon<100$ keV and for the high-energy hump, which is less prominent in unstable regime. These observations comply with the hypothesis \cite{Skalyga_2freq_stabilization} that the secondary frequency enhances losses of high-energy electrons, which in turn leads to a decrease of cyclotron instability growth rate, thus stabilizing the plasma. The given example serves to demonstrate that the two-frequency heating affects the EED whereas the exact interaction mechanism will be discussed in a dedicated follow-up paper.

\section{Plasma stabilization by off-resonance two-frequency heating}

\begin{figure}
%/media/NAS_public/EXPERIMENTS/JYFL2019/2019.02.11/scan3/plotEEDF.m
\includegraphics[width=\linewidth]{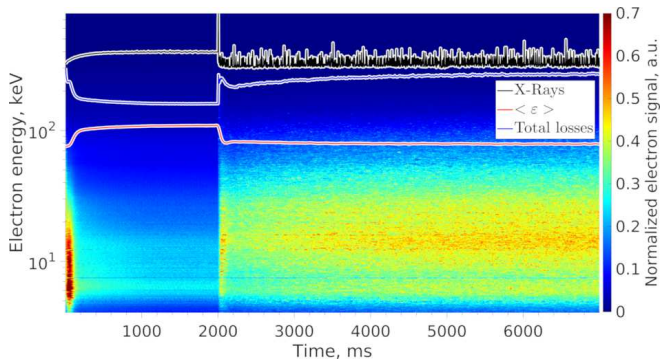}
\caption{\label{fig:fig10} LEED evolution during the transition between stable (double frequency off-resonance heating) and unstable (single frequency heating) plasma. Source settings: 250 W at 14 GHz + 100 W at 11.56 GHz (2 s) / 250 W at 14 GHz (7 s), $B_{min}/B_{ECR}=0.86/1.04$ (calculated for 14 GHz and 11.56 GHz respectively) and $3.5 \cdot 10^{-7}$ mbar oxygen pressure.}
\end{figure}

The hypothesis on the mechanism of the two-frequency plasma stabilization mentioned in the previous section and discussed in detail in \cite{Skalyga_2freq_stabilization} is further supported by the following observations. Figure~\ref{fig:fig10} shows the temporal evolution of the LEED, acquired during periodic power modulation from 100 W (2 s) to 0 W (7 s) at 11.56 GHz on top of constant power of 250 W at 14 GHz, resulting in stable plasma in two-frequency heating regime and unstable plasma in 14 GHz-only regime. The source parameters were the following: $B_{min}/B_{ECR}=0.86/1.04$ (calculated for 14 GHz and 12.7 GHz respectively) and $3.5 \cdot 10^{-7}$ mbar oxygen pressure. It is underlined that the $B_{min}/B_{ECR}>1$ for the secondary frequency means that there was no “cold electron” resonance (i.e. the ECR condition for electrons with $\sim$zero velocity was not met anywhere inside the trap) for the secondary frequency. This implies that the secondary frequency could interact only with electrons with either considerable longitudinal velocity or total kinetic energy, as described by the relativistic Doppler-shifted resonance condition $\omega=\gamma \omega_c (1 \pm n_\parallel \beta_\parallel)$, where $\gamma$~is the relativistic Lorenz factor, $\omega_c$ is the angular cyclotron frequency, $n_\parallel$ is the longitudinal refractive index and $\beta_\parallel = V_\parallel/c$ is the longitudinal electron velocity normalized to the speed of light.

\begin{figure}
%/media/NAS_public/EXPERIMENTS/JYFL2019/2019.02.11/scan3/plotEEDFs.m
\includegraphics[width=\linewidth]{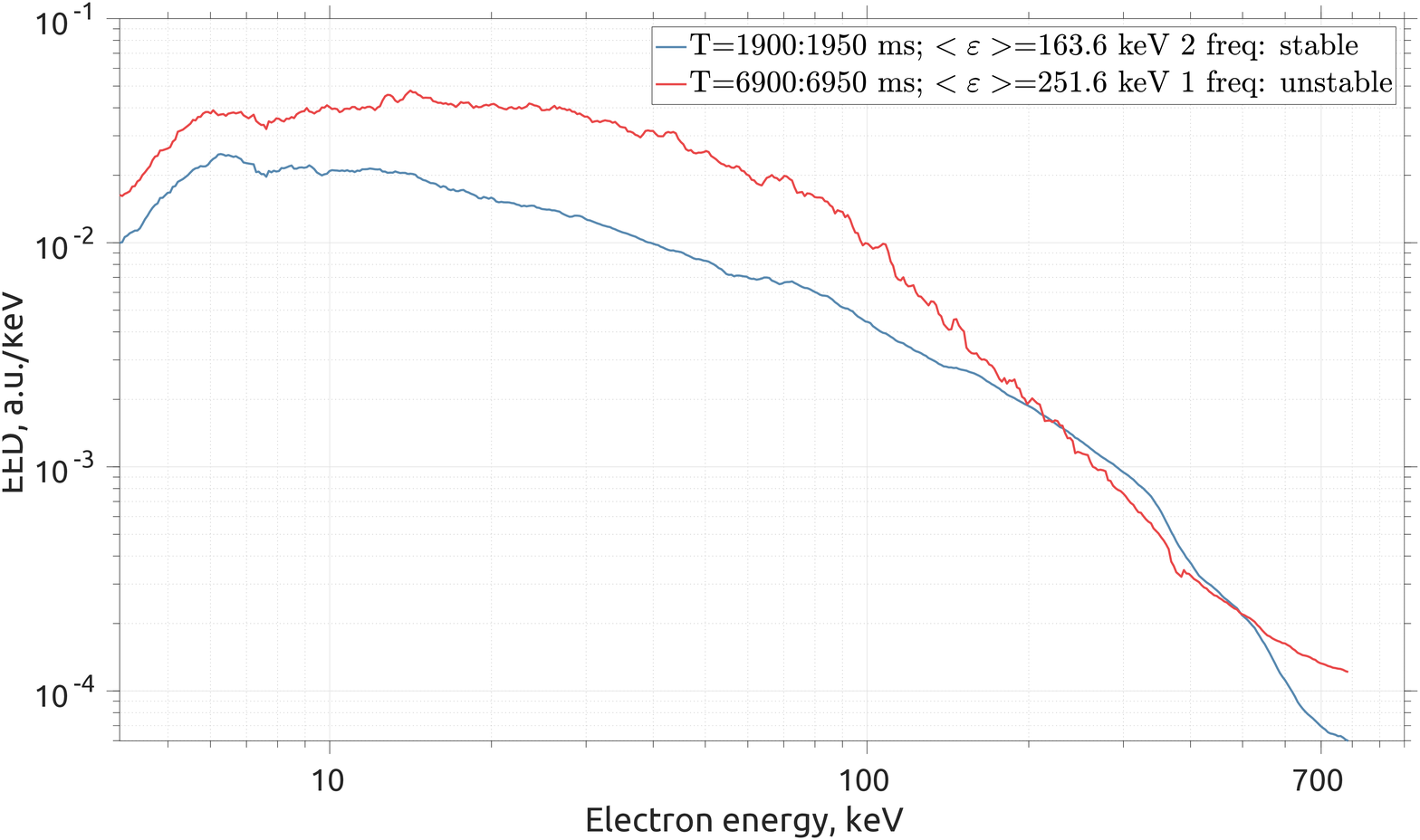}
\caption{\label{fig:fig11} LEED in stable regime, stabilized by off-resonance 2-frequency heating, and unstable plasma. Source settings: same as in Fig.~\ref{fig:fig10}}
\end{figure}

The total losses are noticeably lower in two-frequency, i.e. stable, regime when compared to the single frequency unstable one, contrary to the case of the secondary frequency being above the ``cold” ECR (see Figure~\ref{fig:fig8}). The difference in LEED’s shape (see Figure~\ref{fig:fig8}) is pronounced in the energy range of $\varepsilon<200$ keV, implying that more particles are expelled in the unstable regime in this energy range when compared to the stable regime. With the given settings the high-energy hump has almost vanished in both unstable and stable regimes. 

\section{Discussion}

\begin{figure}
%/media/NAS_public/EXPERIMENTS/JYFL2018/mironov/plotCompareEED_LEED.m
\includegraphics[width=\linewidth]{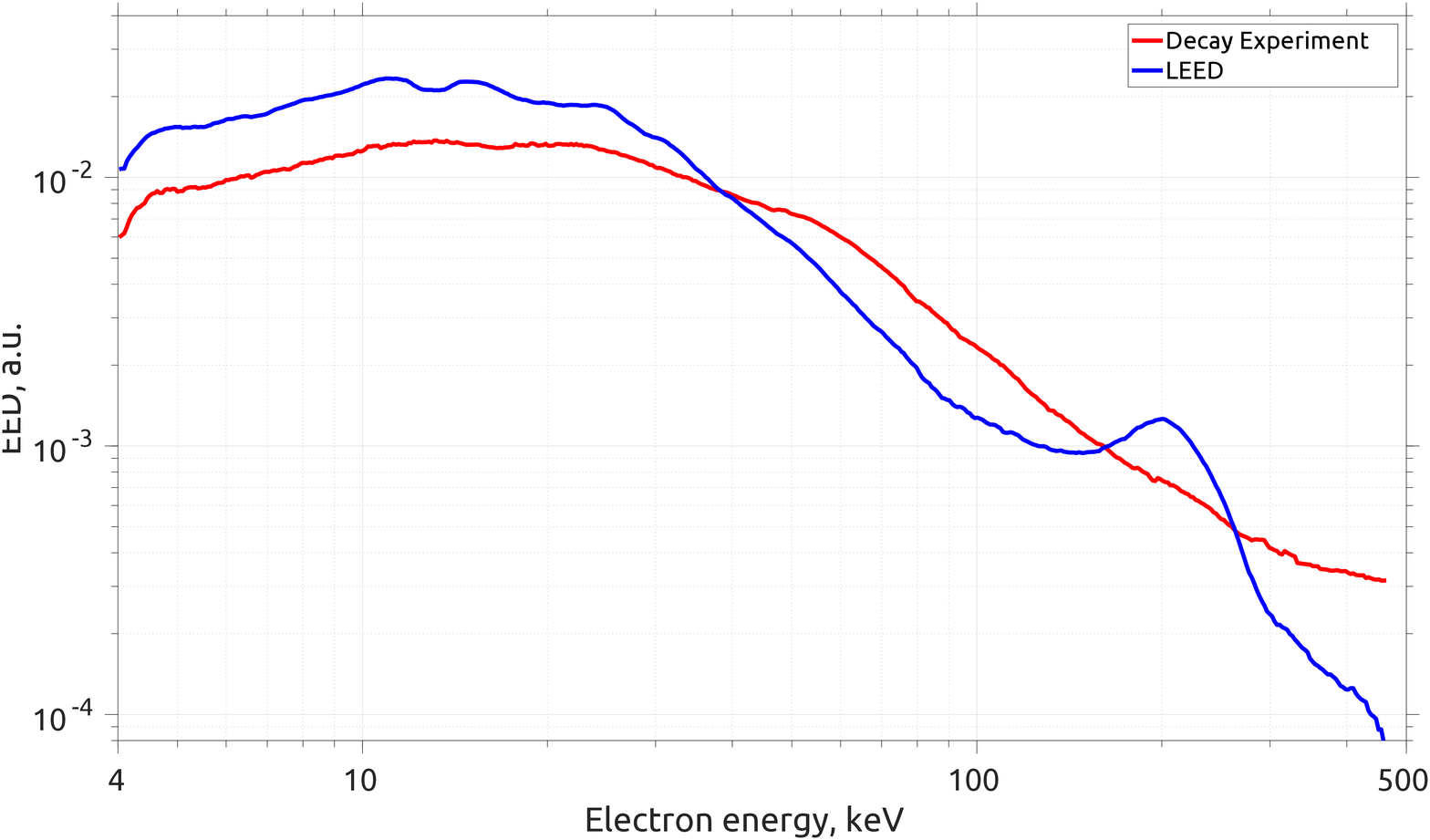}
\caption{\label{fig:fig12} Comparison of the LEED and EEDF reconstructed from the data obtained during the plasma decay. Source settings: single-frequency heating at 14 GHz, 260 W, $B_{min}/B_{ECR}=0.73$, oxygen pressure $3.5 \cdot 10^{-7}$ mbar oxygen pressure.}
\end{figure}

Results obtained with a well-developed technique of measuring the energy distribution of electrons escaping axially from the magnetically confined plasma of an ECRIS, and especially comparing the data in stable and unstable regimes, are discussed above. We would like to underline that despite the method being a novel non-invasive plasma diagnostic tool and yielding direct information on electron energies, it does not provide exact information on the electron energy distribution of the electrons confined in the trap. Here we propose an experimental technique which allows estimating the EEDF inside the trap using the LEED measurements and making certain assumptions. The technique is based on pulsing the microwave power, first allowing the plasma to reach a steady-state, then switching off the microwave radiation and measuring the time-resolved EED of the electrons escaping through the extraction mirror during the plasma decay. In the absence of RF pitch angle scattering and kinetic afterglow instabilities, collisional scattering is the main process pushing electrons into the loss cone. Assuming that the majority of the collisions are elastic (the validity of such assumption needs further investigations) and not changing the electron energy, electron losses integrated over the plasma decay can be argued to represent the information on the EED prevailing inside the plasma at the moment when the microwaves are switched off. An example of such evaluation is presented in Figure~\ref{fig:fig12} (red curve) together with LEED measured before the trailing edge of the heating pulse (blue curve). Both curves are normalized to the unity square. The source settings were the following: single-frequency heating at 14 GHz with 260 W, pulse length 624 ms, repetition rate 1 Hz; $B_{min}/B_{ECR}=0.73$, $3.5 \cdot 10^{-7}$ mbar oxygen pressure.

The overall shape of the two distributions is similar. The main difference is the absence of the high-energy hump in the reconstructed EED (which complies with the presumable origin of the hump being RF-scattering) and a more gentle slope of the reconstructed EED in the energy range of 20--200 keV when compared to the LEED.
More experimental studies and, most likely, comparison to PIC simulations are required to confirm the validity of proposed method.

\begin{acknowledgments}
The experiment was supported by the Academy of Finland Project No~315855. The data processing and analysis was supported by the Russian Science Foundation, project \#19-12-00377. 
\end{acknowledgments}

\nocite{*}
\bibliography{ms}% Produces the bibliography via BibTeX.

\end{document}